\documentclass[aps,prx,twocolumn,floatfix,10pt,amssymb,amsfont,amsmath,superscriptaddress,float,nofootinbib,longbibliography,nobalancelastpage]{revtex4-2}

\usepackage{graphicx} % Required for 
\usepackage{amsmath}
\usepackage{amssymb}
\usepackage{braket}
\usepackage[export]{adjustbox}
\usepackage[T3,T1]{fontenc}
\DeclareSymbolFont{tipa}{T3}{cmr}{m}{n}
\DeclareMathAccent{\fermionic}{\mathalpha}{tipa}{16}

\begin{document}

\title{Local Jordan-Wigner transformations on the torus}

\author{Oliver O'Brien}
\affiliation{Department of Applied Mathematics and Theoretical Physics, University of Cambridge,\\ Wilberforce Road, Cambridge, CB3 0WA, United Kingdom}
\author{Laurens Lootens}
\affiliation{Department of Applied Mathematics and Theoretical Physics, University of Cambridge,\\ Wilberforce Road, Cambridge, CB3 0WA, United Kingdom}
\affiliation{Department of Physics and Astronomy, Ghent University, Krijgslaan 281, 9000 Gent, Belgium}
\author{Frank Verstraete}
\affiliation{Department of Applied Mathematics and Theoretical Physics, University of Cambridge,\\ Wilberforce Road, Cambridge, CB3 0WA, United Kingdom}
\affiliation{Department of Physics and Astronomy, Ghent University, Krijgslaan 281, 9000 Gent, Belgium}

%\date{February 2024}

\begin{abstract}
We present a locality preserving unitary mapping from fermions to qubits on a 2D torus whilst accounting for the mapping of topological sectors. Extending the work of Shukla et al. \cite{Shukla_2020}, an explicit intertwiner is constructed in the form of a projected entangled pair operator. By encoding the information about the charge sectors (and if applicable the twisted boundary conditions) in ancillary qubit(s), the intertwiner becomes a  unitary operator which exchanges boundary conditions and charge sectors.
\end{abstract}

\maketitle

\section{Introduction}
Quantum wavefunctions are described on a projective space, and this allows a system of indistinguishable particles to be described by anti-symmetric wavefunctions, a.k.a. fermions. Although this anti-symmetry prohibits the existence of a tensor product structure, the no-signaling principle can be saved by invoking a super\-selection rule which dictates that all physical observables are of even parity. By imposing a 1-dimensional ordering to a collection of fermions,  Jordan and Wigner \cite{jordan1993paulische} demonstrated how to endow the fermions with a tensor product structure. This was done by introducing a duality mapping between the fermions and a system of spin $1/2$ particles (i.e. qubits). However, this original mapping only preserves locality for 1-D systems. The concept of flux attachment \cite{Jain,Lopez,Halperin} suggested the existence of higher-dimensional analogues of this Jordan-Wigner transformation preserving locality in arbitrary dimensions and in the thermodynamic limit;  the first explicit constructions of such a mapping in 2 spatial dimensions was obtained in \cite{bravyi_fermionic_2002,Verstraete_2005,Ball_gauge}. A system of fermions located on the vertices of a square lattice was mapped to a system of qubits living on the edges of that graph. The fact that this construction yields twice as many qubits as fermionic modes was compensated by the fact that the new system exhibits gauge conditions on every plaquette. This work was followed by many other papers ~\cite{seeley_bravyi-kitaev_2012,Tranter:2015nri,hastings2015, Whitfield:2016lhw,Havl_ek_2017,Setia_2019, steudtner_quantum_2019,jiang2019,Backens_2019,Jiang_2020, Derby_2021, Chiew_2023,Miller_2023,chen_equivalence_2023, PhysRevB.109.115149} that improved the qubit-to-fermion ratio and/or the maximal diameter of the Hamiltonian terms defined on the spin space. The modern point of view is that the fermion to qubit mapping is a duality transformation obtained by gauging the global $Z_2$ parity symmetry of the fermions \cite{Chen_2018}; in such a mapping, the fermion matter can then be disentangled by a local fermionic quantum circuit \cite{Zohar_2018}. However, many questions remained open: it was not clear how to deal with higher dimensional tori with periodic and twisted boundary conditions, and how to modify the mappings when dealing with odd charge sectors. In this paper, we show how those problems can be solved by making use of the formalism of graded tensor networks \cite{Bultinck_2017,Cirac_2021} and more specifically of projected entangled pair operators which explicitely encode the  gauge transformations \cite{Haegeman_2015,Shukla_2020}.

This paper is organized as follows. We start by defining the graded projected entangled  pair operator (PEPO) for the fermion to qubit mapping on arbitrary graphs. Subsequently, we revisit the 1-D Jordan-Wigner transformation for a (anti-)periodic ring of fermions in the even and odd charge sectors in terms of matrix product operators (MPOs). Next, we show how this mapping can readily be extended to the 2-dimensional torus, and pay special attention to the intriguing interplay between charge sectors and boundary conditions. Controlling the possible boundary conditions turns out to be extremely relevant for state of the art simulations of e.g. the Hubbard model \cite{xu2023coexistence}.

\section{Construction}
Given an arbitrary graph $G = (V,E)$, we construct our PEPO by placing  fermionic $Z_2$ graded  $(d+1)$-degree vertex tensors  \cite{Bultinck_2017} at every $d$-degree vertex of $G$ and a graded GHZ tensor on every edge of $G$.

\begin{figure}[hbtp]
    \centering
    \includegraphics{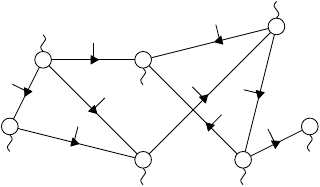}
    \caption{An example PEPO constructed for an arbitary graph with labels and arrows indicating the internal ordering of the vertex tensors and GHZ tensors respectively.}
    \label{fig:example_general_mapping}
\end{figure}

The GHZ tensor is defined as:
\begin{equation}
        \centering
    \includegraphics[trim={0 2 0 0}]{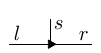} := \sum_{c} \langle c|_s (c|_l (c|_r.
    \label{eq:GHZ}
\end{equation}
where $l$ and $r$ label the left and right virtual fermionic legs of the tensor and $s$ labels the physical spin leg. Every tensor must be drawn with an arrow which dictates what is left and right, but as we work on arbitrary graphs, no spin structure is needed. Curved brackets will be used to denote fermionic degrees of freedom as rounded objects are known to be recalcitrant, while the usual bra-ket notation will be used for the qubit degrees of freedom. We can consider curved brackets to be defined in terms of creation operators acting on the vacuum with $|1) = c^{\dagger}|\Omega), |0) = |\Omega)$ and hence:
\begin{equation}
    |a)|b) = (-1)^{ab}|b)|a),\quad \text{Tr}( |a) (b|) = (-1)^{ab}.
\end{equation}
Throughout this work we refer to the fermionic $\fermionic X$ and $\fermionic Z$ operators which we define as:
\begin{equation}
   \fermionic{X} = \sum_a |a+1) (a|, \quad \fermionic{Z} = \sum_a (-1)^a |a) (a|.
\end{equation}
We study the symmetries given by applying products of these operators to the GHZ tensors: 
\begin{align*}
    \includegraphics[trim={0 2 0 0}]{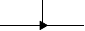} &= \includegraphics[trim={0 2 0 0}]{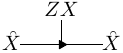} = \sum_{c} \langle c|_s (c|_l (c|_r ZX \fermionic{X}_l \fermionic{X}_r \\
     &=
    \includegraphics[trim={0 2 0 0}]{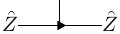} = 
    \includegraphics[trim={0 2 0 0}]{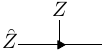}.
\end{align*}
When writing symmetries in this graphical notation, we require that the operators applied are ordered with the same internal ordering of the legs which they apply to. The vertex tensor is defined as:
\begin{equation}
    \centering
        \label{eq:vertex_tensor}
    \includegraphics[valign=c]{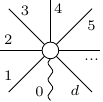} := \hspace{-1em} \sum_{\substack{\{a_i\} \\ \sum_i a_i = 0\, (\text{mod}\, 2)}} \hspace{-1em} |a_d)\ldots |a_1) |a_0),
\end{equation} 
where $a_0$ labels the physical fermionic leg of the tensor and $a_1, \ldots, a_d$ label the virtual fermionic legs which will be contracted with the GHZ tensor; the ordering is important and must be specified on every vertex. In this work we always using the ordering of clockwise starting at the physical fermionic leg. Any choice of orderings will be valid, however, changing the internal ordering will lead to different mappings. With this in mind we consider the symmetries of the vertex tensors:
\begin{equation}
    \includegraphics[valign=c]{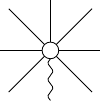} = \includegraphics[valign=c]{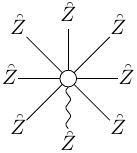}
    = \includegraphics[valign=c]{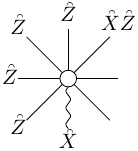}.
\end{equation}
The constructed PEPO (Figure~\ref{fig:example_general_mapping}) will map fermionic operators acting on the vertices of $G$ to spin operators acting on the edges of $G$. We represent this as $\bra{s}(E|  |V_v)|V_p)$ where $\bra{s}$ are the physical spins, $(E|$ are the virtual fermions on the GHZ tensors, $|V_v)$ are the virtual fermions on the vertex tensors and $|V_p)$ are the physical fermions. This mapping will be local in the sense that fermionic hopping terms acting along edges of the graph $G$ will map to spin operators with a maximum Pauli weight $2\Delta-1$ (where $\Delta$ is the maximum degree of $G$).

As all tensors involved in the PEPO have an even parity, a tensor network diagram uniquely specifies the mapping, as long as arrows are put on all edges and an internal ordering of vertex tensors is chosen. Note that defining a fermionic operator (e.g. $\fermionic{O}= a_j^{\dagger} a_i$) requires a linear ordering for the Hilbert space it acts upon (e.g. $j < i$). A change to this ordering (e.g. to $i < j$) incurs a global minus sign on the fermionic operator giving a new definition of the same operator ($\fermionic{O} = -a_j^{\dagger} a_i$) acting on a different Hilbert space. As our mapping is not affected by the choice of linear ordering it incurs no sign change upon reordering (it will always map $a_j^{\dagger} a_i$ to $O$). Therefore, whether $\fermionic{O}$ is mapped to $+ O$ or $- O$ depends on the linear ordering chosen for the fermionic Hilbert space being mapped. 

Using the symmetry properties of our tensors, fermionic operators acting on the physical fermionic degrees of freedom ($O_{pf}$) are transformed to to fermionic operators acting on the virtual fermionic legs of our vertex tensors ($O_{vf (ver)}$). This is equivalent to acting with the same operators on the virtual fermionic legs of our edge tensors ($O_{vf (edg)}$), which then transform them to spin operators acting on the physical spin degrees of freedom ($O_{s}$):
\begin{align*}
    O_{pf}\bra{s}(E|  |V_v)|V_p) &= \bra{s}(E|O_{pf}  |V_v)|V_p)\\
    &= \bra{s}(E| O_{vf (ver)}|V_v)|V_p)\\ 
    &= \bra{s}(E| O_{vf (edg)}|V_v)|V_p)\\
    &= \bra{s}O_s(E| |V_v)|V_p) \\
    &= \bra{s}(E| |V_v)|V_p)O_s.
\end{align*}

The symmetries of our tensors reveal that the theory at the side of the spins is a gauge theory, with a gauge condition for every possible cycle in the graph. The number of independent gauge conditions is given by the circuit rank $r$ of $G$ (the number of independent cycles in $G$). As $r = |E| - |V| + 1$, this mapping is an isomorphism from the even fermionic sector to the $+1$ joint eigenspace of the gauge conditions on the spin side. On a torus two of these independent gauge conditions correspond to non-contractible loops; these gauge conditions can be lifted and interpreted as charge sectors, and those different charge sectors will correspond to (twisted) boundary conditions on the fermion side.

% \begin{figure}[hbtp]
%     \centering
% \includegraphics[page=4]{IPE_figs.pdf}
% \caption{Symmetries of the GHZ tensor. The operators labeled are all applied to the right of the tensor and in the order specified. For example, the last diagram corresponds to: $\sum_{c} \bra{c}_S(c|_{L}(c|_{R} = \sum_{c} \bra{c}_S(c|_{L}(c|_{R} Z_S X_S X_L X_R$}
%     \label{fig:GHZ_tensor_symmetries}
% \end{figure}
% \begin{figure}[hbtp]
%     \centering
%     \includegraphics[page=8]{IPE_figs.pdf}
%     \caption{Symmetries of the general vertex tensor. The operators labeled are applied to the left of the tensor and in the order specified. }
%     \label{fig:vertex_tensor_symmetries}
% \end{figure}

To admit a mapping of odd fermionic operators we add a single odd tensor defect (fermionic $\fermionic{X}$ operator) to a virtual leg of the PEPO. This can be achieved by acting with a single $C\fermionic{X}$ gate
\begin{equation}
    C\fermionic{X} = \sum_{a,c} |a+c) (a| \bra{c}
\end{equation}
to an edge of our PEPO. This fermionic charge defect can be placed anywhere, and its location can be understood as a boundary condition for the dual spin theory: a manifestation of the effect that dualities switch boundary conditions with charge sectors.

%As this PEPO is a unitary map there must be $|E|+1-|V|$ independent gauge conditions which arise from all possible cycles in $G$. We may violate a gauge condition by adding a boundary condition and considering the resulting PEPO as a map into a non-trivial charge sector of this gauge condition. Equivalently, we can consider the fermionic $X$ operator as the boundary condition mapping into the odd parity fermionic charge sector.

A single fermionic creation/annihilation operator will be mapped to a string of spins terminating at the odd defect. If strings of two different operators converge at any point before reaching the defect they will anticommute. If two strings never converge but instead approach the defect from opposite directions, then they will necessarily anticommute on the closest spin to the defect:
\begin{equation}
\includegraphics[valign=c]{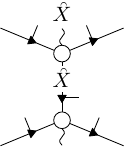} = 
\includegraphics[valign=c]{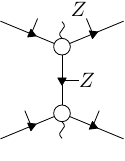},
\end{equation}
\begin{equation}
\includegraphics[valign=c]{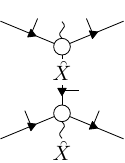} = 
\includegraphics[valign=c]{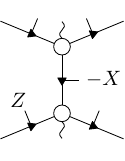}.
\end{equation}
Hence, we produce the correct anti-commutation relations for the fermionic algebra.

Due to the equivalence of 2D fermionic encodings shown by \cite{chen_equivalence_2023} it is possible to reconstruct any existing 2D fermion-to-qubit mapping through a particular choice of $G$, internal orderings, single qubit gates and majorana repairing. Further the inclusion of the $\fermionic X$ defect allows for the representation of tree-based mappings such as the Bravyi-Kitaev Fast Mapping \cite{bravyi_fermionic_2002}. We will expand on this equivalence in future work.

\section{1D construction}
\label{section:1d-construction}
On a 1D cycle graph, our construction turns out to be equivalent to the Jordan-Wigner mapping composed with the Kramers–Wannier duality. We will use this as an example to illustrate our formalism.

The 1D PEPO is an MPO formed of interleaved GHZ tensors and 3-degree vertex tensors. The mappings of fermionic $\fermionic{Z}$ and $\fermionic{X}\fermionic{X}$ operations can be inferred from the invariance of the MPO under the following operations:
\begin{equation*}
    \centering
    \includegraphics[valign=c]{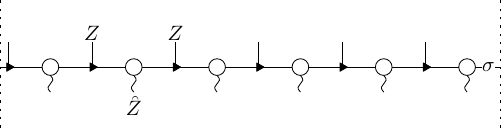}.
    \end{equation*}
    \begin{equation*}
    \includegraphics[valign=c]{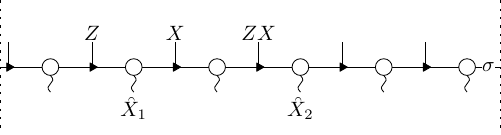}\,.
    \label{fig:1D-JW-mapping}
\end{equation*}
In this particular example (with $\sigma$ boundary condition), a fermionic $\fermionic{Z_i}$ operator is mapped to a $Z_{i-1/2}Z_{i+1/2}$ term on the spin side, and a fermionic $\fermionic{X}_{i-1}I_i\fermionic{X}_{i+1}$ to $Z_{i-3/2}X_{i-1/2}(XZ)_{i+1/2}$. Different choices of boundary condition will map different topological sectors. These sectors are characterised by the charge sector under the global symmetries of each side, and the twists required for translational symmetries. The charge sector mapped by a boundary condition is determined by the sign produced when the MPO absorbs the given symmetry. As in the example below, it is important to account for the signs due to moving any unpaired fermionic operators at the boundaries next to each other and into the correct order to be absorbed by the vertex symmetries without additional signs.
% \begin{align*}
%     \centering
%     &\includegraphics[page=17, valign=c]{IPE_figs.pdf}\\
%     &\hskip 12em
%         \scalebox{1.5}{\rotatebox{90}{$\,=$}}\\
%     &\includegraphics[page=18, valign=c]{IPE_figs.pdf}\\
%     &\hskip 12em
%         \scalebox{1.5}{\rotatebox{90}{$\,=$}}\\
%     &\includegraphics[page=19, valign=c]{IPE_figs.pdf}
% \end{align*}
\begin{align*}
    &\bra{s}(E|  \sigma_n|V_v)|V_p) \prod_i X_i = \bra{s}\prod_i X_i(E|  \sigma_n|V_v)|V_p)\\
    &= \bra{s}(E|Y_1X_2Y_3 \ldots X_{n-2}Y_{n-1} X_n \sigma_n|V_v)|V_p) \\
    &= \bra{s}(E|Y_1 X_n \sigma_n|V_v)|V_p) \\
    &= \begin{cases}-\bra{s}(E|X_n \sigma_n Y_1 |V_v)|V_p) & \sigma \text{ is even}\\\bra{s}(E|X_n \sigma_n Y_1 |V_v)|V_p) & \sigma \text{ is odd}\end{cases} \\
    &= \begin{cases} -\bra{s}(E| \sigma_n |V_v)|V_p) & \sigma =I\\\bra{s}(E|  \sigma_n|V_v)|V_p) & \sigma =Z\\\bra{s}(E| \sigma_n |V_v)|V_p) & \sigma =X\\
    -\bra{s}(E| \sigma_n |V_v)|V_p) & \sigma =Y \end{cases}
\end{align*}
The twist produced by a boundary condition under a translation is calculated by shifting over each physical leg above and below the MPO, and considering the necessary operators produced on the physical legs by moving the boundary condition to the new boundary. For example, translating the $\fermionic Z$ boundary condition produces a $\fermionic Z$ twist on a fermionic physical leg:
\begin{align*}
    \centering
    &\includegraphics[valign=c]{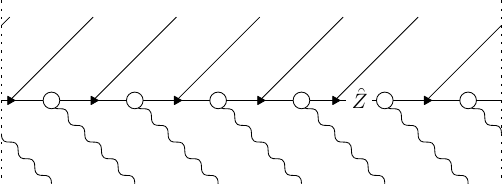}\\
    &\hskip 12em
        \rotatebox{90}{$\,=$}\\
    &\includegraphics[valign=c]{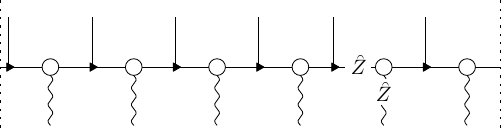}.
\end{align*}
Table~\ref{tab:cycle_mapping} illustrates how the topological sectors of the 1D cyclical graph are determined and mapped by different boundary conditions. By treating these boundary conditions as additional quantum degrees of freedom, we are able to recover a completely unitary transformation from fermions to spins. If we did not consider boundary conditions we would only have an isometry between the periodic even fermionic space and the periodic $-1$ spin eigenspace of $\prod_i X_i$, rather than mapping the entire fermionic Hilbert space to the entire spin Hilbert space.
\begin{table}[htbp]
    \begin{tabular}{|c c| c|c c |c| c c |c|}
    \hline
    \multicolumn{2}{|c|}{Fermions}& & \multicolumn{2}{|c|}{Spins} & &\multicolumn{2}{|c|}{Dual Spins}& Unified  \\
    $\prod_i \fermionic{Z}_i$ & Twist & BC 1& $\prod_i X_i$ & Twist & BC 2 & $\prod_i Z_i$ & Twist & BC\\
    \hline
    $+1$ & $I$ & $I$&$- 1$ & $I$ &$Z$ &$+1$ & $Z$ & $\fermionic{Z}$\\
    $+1$ & $\fermionic{Z}$ &$\fermionic Z$ &$+ 1$ & $I$ & $I$&$+1$  & $I$& $\fermionic{Z}$\\
    $- 1$ & $I$&$\fermionic X$& $+1$ & $X$  &$X$ &$-1$ & $I$& $\fermionic{Z}\fermionic{X}$\\
    $-1$ & $\fermionic{Z}$ &$\fermionic{Z}\fermionic X$&$- 1$ & $X$ &$ZX$&$-1$ & $Z$ & $\fermionic{Z}\fermionic{X}$\\
    \hline
    \end{tabular}
    \caption{Mapping of topological sectors of fermions to spins on the chain dependent on the choice of boundary condition (BC 1) for the duality MPO. These sectors can then be mapped again to the dual spins by the application of a Kramers-Wannier MPO with a suitable choice of boundary condition (BC 2). The composed mapping from fermions to dual spins is equivalent to the canonical Jordan-Wigner mapping with one single boundary condition (Unified BC). \label{tab:cycle_mapping}}
\end{table}

We may compose our MPO with the Kramers-Wannier MPO in order to recover the canonical Jordan-Wigner transformation. The composition will necessarily evaluate to zero if the charge sectors are not matched. However, mismatched twists will give a non-zero MPO, therefore, it is necessary to chose the unique KW boundary condition which matches the twist and charge sector as seen in Table~\ref{tab:cycle_mapping}. We can then combine these two boundary conditions as follows:
\begin{equation*}
     \includegraphics[valign=c]{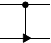} =\includegraphics[valign=c]{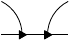} \implies \includegraphics[valign=c]{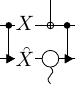} = \includegraphics[valign=c]{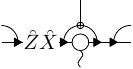}\,.
\end{equation*}
Hence, the $\fermionic Z$ and $\fermionic Z\fermionic X$ boundary conditions map both the periodic and anti-periodic boundary conditions for the even and odd sectors respectively. The $I$ and $\fermionic X$  boundary conditions evaluate to 0 (as they correspond to mismatched charge sectors between the combined MPOs), so we can simply add them to their non-zero counterparts ($\fermionic Z$ and $\fermionic Z\fermionic X$) to obtain rank 1 boundary conditions $|0)(0|$ and $|1)(0|$. By summing these two MPOs we produce a unified MPO with boundary $\big(|0)+|1)\big)(0|$ giving a unitary mapping from fermions to spins. This clarifies the mystery of why one and the same JW-mapping can deal with all four topological sectors - which is very unusual and unexpected for a duality transformation. 

We note that using our formalism the boundary conditions are sensitive to their placement in the MPO. For example, in 1D the $\fermionic{Y}$ boundary condition to the right of a 3-degree vertex tensor is similar to a $\fermionic{X}$ boundary condition to the left of a 3-degree vertex tensor. For consistency we always refer to the boundary condition as acting to the left of the 3-degree vertex tensor.
\section{2D construction}
\label{section:2d-construction}
For any planar graph $G$ on an orientable manifold, there will be a natural choice for the orientations of the vertex tensors. In this section we exemplify our mapping using the square lattice. In the bulk, our PEPO is similar to the one constructed in \cite{Shukla_2020} and consists out of two simple tensors, a GHZ tensor and a 5-degree vertex tensor, arranged on the edges and vertices of a square lattice. On every plaquette, the symmetries impose that the spin system has to satisfy a gauge condition:
\begin{equation}
    \includegraphics[valign=c]{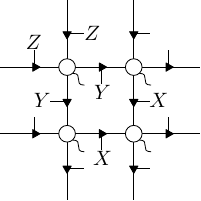} = \includegraphics[valign=c]{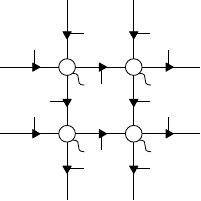}.
\end{equation}
The number of vertices is exactly equal to the number of edges minus the number of plaquettes, this is the complete set of gauge conditions, if it were not for the fact that two of them are not independent. Of course, the two missing gauge conditions are the ones winding around the torus. On a periodic torus we thus have two non-contractible independent gauge conditions. These are given by loops of spin operators along the horizontal ($X_H$, magenta) and vertical ($X_V$, green) directions:
\begin{equation}
    \includegraphics[valign=c]{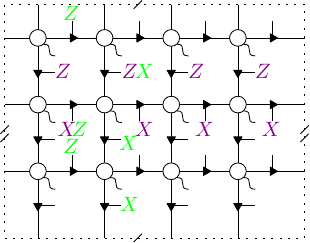}.
\end{equation}
From the point of view of the fermion side, changing the charge sector on the spin side is equivalent to adding twisted boundary conditions.  These boundary conditions are strings of fermionic $\fermionic Z$ operators lying along the virtual legs of the PEPO in the horizontal ($\fermionic{Z}_H$) and vertical ($\fermionic{Z}_V$) directions respectively. Table~\ref{tab:torus_mapping} illustrates how these boundary conditions are mapped on a torus. It is quite marvellous to see how the four boundary conditions and two charge sectors are mapped to two boundary conditions and four charge sectors. Boundary conditions are sensitive to their position in the tensor network, so for simplicity we will only place the $X$ defect on horizontal virtual legs to the right of the GHZ tensor. There is a freedom of choice for which non-contractible loops to choose as $X_H$ and $X_V$. We have chosen them such that they intersect with the $\fermionic X$ defect if present, so the table matches the results for the 1D case. However, it would be equally valid to choose them to avoid the $\fermionic X$ defect which would make the spin charge sectors symmetrical for the even and odd fermionic sectors.

Below is an example of a PEPO with a $\fermionic Z_H$ boundary condition and an $\fermionic X$ defect, which will map between the odd fermionic algebra and the $(-1,+1)$ eigenspace of $( X_H, X_V)$:
\begin{equation}
     \includegraphics[valign=c]{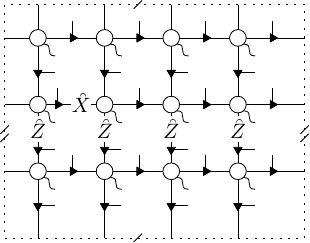}.
\end{equation}

% We consider the even fermionic algebra in terms of vertex $V_i$ and hopping $E_{ij}$ operators:
% \begin{equation}
% V_i = Z_i
% \end{equation}
% \begin{equation}
% E_{ij} =  X_i  X_j
% \end{equation}
% where $X_i$ and $Z_i$ are fermionic $\sigma_x$ and $\sigma_z$ respectively.

% This basis along with conditions \eqref{eq:anticommutation_relations} and \eqref{eq:loop_condition} completely defines the even fermionic algebra, hence our aim is to produce spin representations of $V_i$ and $E_{ij}$ which satisfy the following anticommutation relations
%     \begin{align}\begin{aligned}
%     \label{eq:anticommutation_relations}
%                 \{ E_{jk}, V_j\} = 0, \{ E_{ij}, E_{jk} \} &= 0 \> \forall \>i \neq k\\
%                 [V_i, V_j] =0, [E_{ij}, V_m] = 0, [E_{ij}, E_{mn}] &= 0 \>\forall \>i \neq j \neq m \neq n
% \end{aligned}
% \end{align}
% along with the following loop condition for any closed loop of sites $p_i$:
% \begin{equation}
%     \label{eq:loop_condition}
%         i^{n} \prod_{i=1}^{n} E_{p_i p_{i+1}} = 1
% \end{equation}
% By applying our PEPO to $V_i$ and $E_{ij}$ we can find spin representations, which can then be verified by checking conditions \eqref{eq:anticommutation_relations} and \eqref{eq:loop_condition}.
\begin{table}
    \centering
    \begin{tabular}{|c c c| c| c c c |}
    \hline
    \multicolumn{3}{|c|}{Fermions} && \multicolumn{3}{|c|}{Spins}  \\
    $\prod_i \fermionic{Z}_i$ & H. Twist& V. Twist & BC & $X_H$ & $X_V$&H./V. Twist\\
    \hline
    $+1$ & $I$&$I$ &$I$ &$-1$ &$-1$&$I$\\
    $+1$ & $I$&$\fermionic{Z}_{p_H}$  &$\fermionic{Z}_H$ &$-1$ & $+1$&$I$\\
    $+1$ & $\fermionic{Z}_{p_V}$&$I$ &$\fermionic{Z}_V$ &$+1$ &$-1$&$I$\\
    $+1$ & $\fermionic{Z}_{p_V}$&$\fermionic{Z}_{p_H}$ &$\fermionic{Z}_V\fermionic{Z}_H$ &$+1$ &$+1$&$I$\\
    $-1$ &$I$ & $I$&$\fermionic{X}$ &$+1$ & $+1$&$Z_vX_h/Z_hX_v$\\
    $-1$ &$I$ &$\fermionic{Z}_{p_H}$ &$\fermionic{Z}_H\fermionic X$ &$+1$ &$-1$&$Z_vX_h/Z_hX_v$\\
    $-1$ & $\fermionic{Z}_{p_V}$&$I$&$\fermionic{Z}_V\fermionic X$ &$-1$ &$+1$&$Z_vX_h/Z_hX_v$\\
    $-1$ & $\fermionic{Z}_{p_V}$& $\fermionic{Z}_{p_H}$&$\fermionic{Z}_H\fermionic{Z}_V\fermionic X$ &$-1$ &$-1$&$Z_vX_h/Z_hX_v$\\
    \hline
    \end{tabular}
    \caption{Table illustrating how different boundary conditions of the PEPO on the torus map different charge sectors and produce different twists. The $\fermionic Z_{p_V}$ and $\fermionic Z_{p_H}$ twists are vertical and horizontal loops of $\fermionic Z$ operators on the physical fermions. The $Z_vX_h/Z_hX_v$ twist (with subscript $v/h$ indicating whether the operator is applied on a vertical or horizontal leg) constitutes a single unit cell of $X_H$ or $X_V$ depending on whether the twist was taken in the horizontal or vertical direction respectively.  \label{tab:torus_mapping}
}
\end{table}
\section{Conclusion}
We have shown how boundary conditions on a torus relate to various topological charge sectors. By treating those boundary degrees of freedom as extra quantum degrees of freedom, this allows us to construct a unitary operator (as opposed to an isometry) 
implementing the duality transform, demonstrating that the higher dimensional JW transformation completely preserves the spectrum of this bigger theory. This is completely in line with the results in \cite{lootens2023lowdepth, Lootens_2023, Lootens:2022avn} for all possible dualities defined in terms of categorical symmetries. 

Further, we have demonstrated a tensor network approach to constructing arbitrary fermion to qubit encodings. The resulting spin system is a gauge theory with a very particular type of gauge constraints. We conjecture that such gauge theories in two dimensions can only be ``ungauged''  with graded tensor network intertwiners. This highlights the difference between the 1- and the 2-D case, as fermions seem to be indispenable in the latter.

\medskip\noindent
\textbf{Acknowledgments:} \emph{We thank Clement Delcamp for helpful discussions and collaboration on \cite{lootens2023lowdepth,Lootens_2023, Lootens:2022avn}. This work has received funding from EOS (grant No. 40007526), IBOF (grant No. IBOF23/064), and BOF-GOA (grant No. BOF23/GOA/021). LL is supported by a postdoctoral grant 12AUN24N from the Research Foundation Flanders (FWO).}

\newpage

\bibliography{references.bib}

\begin{thebibliography}{32}%
\makeatletter
\providecommand \@ifxundefined [1]{%
 \@ifx{#1\undefined}
}%
\providecommand \@ifnum [1]{%
 \ifnum #1\expandafter \@firstoftwo
 \else \expandafter \@secondoftwo
 \fi
}%
\providecommand \@ifx [1]{%
 \ifx #1\expandafter \@firstoftwo
 \else \expandafter \@secondoftwo
 \fi
}%
\providecommand \natexlab [1]{#1}%
\providecommand \enquote  [1]{``#1''}%
\providecommand \bibnamefont  [1]{#1}%
\providecommand \bibfnamefont [1]{#1}%
\providecommand \citenamefont [1]{#1}%
\providecommand \href@noop [0]{\@secondoftwo}%
\providecommand \href [0]{\begingroup \@sanitize@url \@href}%
\providecommand \@href[1]{\@@startlink{#1}\@@href}%
\providecommand \@@href[1]{\endgroup#1\@@endlink}%
\providecommand \@sanitize@url [0]{\catcode `\\12\catcode `\$12\catcode `\&12\catcode `\#12\catcode `\^12\catcode `\_12\catcode `\%12\relax}%
\providecommand \@@startlink[1]{}%
\providecommand \@@endlink[0]{}%
\providecommand \url  [0]{\begingroup\@sanitize@url \@url }%
\providecommand \@url [1]{\endgroup\@href {#1}{\urlprefix }}%
\providecommand \urlprefix  [0]{URL }%
\providecommand \Eprint [0]{\href }%
\providecommand \doibase [0]{https://doi.org/}%
\providecommand \selectlanguage [0]{\@gobble}%
\providecommand \bibinfo  [0]{\@secondoftwo}%
\providecommand \bibfield  [0]{\@secondoftwo}%
\providecommand \translation [1]{[#1]}%
\providecommand \BibitemOpen [0]{}%
\providecommand \bibitemStop [0]{}%
\providecommand \bibitemNoStop [0]{.\EOS\space}%
\providecommand \EOS [0]{\spacefactor3000\relax}%
\providecommand \BibitemShut  [1]{\csname bibitem#1\endcsname}%
\let\auto@bib@innerbib\@empty
%</preamble>
\bibitem [{\citenamefont {Shukla}\ \emph {et~al.}(2020)\citenamefont {Shukla}, \citenamefont {Ellison},\ and\ \citenamefont {Fidkowski}}]{Shukla_2020}%
  \BibitemOpen
  \bibfield  {author} {\bibinfo {author} {\bibfnamefont {S.~K.}\ \bibnamefont {Shukla}}, \bibinfo {author} {\bibfnamefont {T.~D.}\ \bibnamefont {Ellison}},\ and\ \bibinfo {author} {\bibfnamefont {L.}~\bibnamefont {Fidkowski}},\ }\bibfield  {title} {\bibinfo {title} {Tensor network approach to two-dimensional bosonization},\ }\bibfield  {journal} {\bibinfo  {journal} {Physical Review B}\ }\textbf {\bibinfo {volume} {101}},\ \href {https://doi.org/10.1103/physrevb.101.155105} {10.1103/physrevb.101.155105} (\bibinfo {year} {2020})\BibitemShut {NoStop}%
\bibitem [{\citenamefont {Jordan}\ and\ \citenamefont {Wigner}(1993)}]{jordan1993paulische}%
  \BibitemOpen
  \bibfield  {author} {\bibinfo {author} {\bibfnamefont {P.}~\bibnamefont {Jordan}}\ and\ \bibinfo {author} {\bibfnamefont {E.~P.}\ \bibnamefont {Wigner}},\ }\bibfield  {title} {\bibinfo {title} {{\"U}ber das paulische {\"a}quivalenzverbot},\ }in\ \href@noop {} {\emph {\bibinfo {booktitle} {The Collected Works of Eugene Paul Wigner}}}\ (\bibinfo  {publisher} {Springer},\ \bibinfo {year} {1993})\ pp.\ \bibinfo {pages} {109--129}\BibitemShut {NoStop}%
\bibitem [{\citenamefont {Jain}(1989)}]{Jain}%
  \BibitemOpen
  \bibfield  {author} {\bibinfo {author} {\bibfnamefont {J.~K.}\ \bibnamefont {Jain}},\ }\bibfield  {title} {\bibinfo {title} {Incompressible quantum hall states},\ }\href {https://doi.org/10.1103/PhysRevB.40.8079} {\bibfield  {journal} {\bibinfo  {journal} {Phys. Rev. B}\ }\textbf {\bibinfo {volume} {40}},\ \bibinfo {pages} {8079} (\bibinfo {year} {1989})}\BibitemShut {NoStop}%
\bibitem [{\citenamefont {Lopez}\ and\ \citenamefont {Fradkin}(1991)}]{Lopez}%
  \BibitemOpen
  \bibfield  {author} {\bibinfo {author} {\bibfnamefont {A.}~\bibnamefont {Lopez}}\ and\ \bibinfo {author} {\bibfnamefont {E.}~\bibnamefont {Fradkin}},\ }\bibfield  {title} {\bibinfo {title} {Fractional quantum hall effect and chern-simons gauge theories},\ }\href {https://doi.org/10.1103/PhysRevB.44.5246} {\bibfield  {journal} {\bibinfo  {journal} {Phys. Rev. B}\ }\textbf {\bibinfo {volume} {44}},\ \bibinfo {pages} {5246} (\bibinfo {year} {1991})}\BibitemShut {NoStop}%
\bibitem [{\citenamefont {Halperin}\ \emph {et~al.}(1993)\citenamefont {Halperin}, \citenamefont {Lee},\ and\ \citenamefont {Read}}]{Halperin}%
  \BibitemOpen
  \bibfield  {author} {\bibinfo {author} {\bibfnamefont {B.~I.}\ \bibnamefont {Halperin}}, \bibinfo {author} {\bibfnamefont {P.~A.}\ \bibnamefont {Lee}},\ and\ \bibinfo {author} {\bibfnamefont {N.}~\bibnamefont {Read}},\ }\bibfield  {title} {\bibinfo {title} {Theory of the half-filled landau level},\ }\href {https://doi.org/10.1103/PhysRevB.47.7312} {\bibfield  {journal} {\bibinfo  {journal} {Phys. Rev. B}\ }\textbf {\bibinfo {volume} {47}},\ \bibinfo {pages} {7312} (\bibinfo {year} {1993})}\BibitemShut {NoStop}%
\bibitem [{\citenamefont {Bravyi}\ and\ \citenamefont {Kitaev}(2002)}]{bravyi_fermionic_2002}%
  \BibitemOpen
  \bibfield  {author} {\bibinfo {author} {\bibfnamefont {S.}~\bibnamefont {Bravyi}}\ and\ \bibinfo {author} {\bibfnamefont {A.}~\bibnamefont {Kitaev}},\ }\bibfield  {title} {\bibinfo {title} {Fermionic quantum computation},\ }\href {https://doi.org/10.1006/aphy.2002.6254} {\bibfield  {journal} {\bibinfo  {journal} {Annals of Physics}\ }\textbf {\bibinfo {volume} {298}},\ \bibinfo {pages} {210} (\bibinfo {year} {2002})},\ \bibinfo {note} {arXiv: quant-ph/0003137}\BibitemShut {NoStop}%
\bibitem [{\citenamefont {Verstraete}\ and\ \citenamefont {Cirac}(2005)}]{Verstraete_2005}%
  \BibitemOpen
  \bibfield  {author} {\bibinfo {author} {\bibfnamefont {F.}~\bibnamefont {Verstraete}}\ and\ \bibinfo {author} {\bibfnamefont {J.~I.}\ \bibnamefont {Cirac}},\ }\bibfield  {title} {\bibinfo {title} {Mapping local hamiltonians of fermions to local hamiltonians of spins},\ }\href {https://doi.org/10.1088/1742-5468/2005/09/p09012} {\bibfield  {journal} {\bibinfo  {journal} {Journal of Statistical Mechanics: Theory and Experiment}\ }\textbf {\bibinfo {volume} {2005}},\ \bibinfo {pages} {P09012–P09012} (\bibinfo {year} {2005})}\BibitemShut {NoStop}%
\bibitem [{\citenamefont {Ball}(2005)}]{Ball_gauge}%
  \BibitemOpen
  \bibfield  {author} {\bibinfo {author} {\bibfnamefont {R.~C.}\ \bibnamefont {Ball}},\ }\bibfield  {title} {\bibinfo {title} {Fermions without fermion fields},\ }\href {https://doi.org/10.1103/PhysRevLett.95.176407} {\bibfield  {journal} {\bibinfo  {journal} {Phys. Rev. Lett.}\ }\textbf {\bibinfo {volume} {95}},\ \bibinfo {pages} {176407} (\bibinfo {year} {2005})}\BibitemShut {NoStop}%
\bibitem [{\citenamefont {Seeley}\ \emph {et~al.}(2012)\citenamefont {Seeley}, \citenamefont {Richard},\ and\ \citenamefont {Love}}]{seeley_bravyi-kitaev_2012}%
  \BibitemOpen
  \bibfield  {author} {\bibinfo {author} {\bibfnamefont {J.~T.}\ \bibnamefont {Seeley}}, \bibinfo {author} {\bibfnamefont {M.~J.}\ \bibnamefont {Richard}},\ and\ \bibinfo {author} {\bibfnamefont {P.~J.}\ \bibnamefont {Love}},\ }\bibfield  {title} {\bibinfo {title} {The {Bravyi}-{Kitaev} transformation for quantum computation of electronic structure},\ }\href {https://doi.org/10.1063/1.4768229} {\bibfield  {journal} {\bibinfo  {journal} {The Journal of Chemical Physics}\ }\textbf {\bibinfo {volume} {137}},\ \bibinfo {pages} {224109} (\bibinfo {year} {2012})},\ \bibinfo {note} {arXiv: 1208.5986}\BibitemShut {NoStop}%
\bibitem [{\citenamefont {Tranter}\ \emph {et~al.}(2015)\citenamefont {Tranter}, \citenamefont {Sofia}, \citenamefont {Seeley}, \citenamefont {Kaicher}, \citenamefont {McClean}, \citenamefont {Babbush}, \citenamefont {Coveney}, \citenamefont {Mintert}, \citenamefont {Wilhelm},\ and\ \citenamefont {Love}}]{Tranter:2015nri}%
  \BibitemOpen
  \bibfield  {author} {\bibinfo {author} {\bibfnamefont {A.}~\bibnamefont {Tranter}}, \bibinfo {author} {\bibfnamefont {S.}~\bibnamefont {Sofia}}, \bibinfo {author} {\bibfnamefont {J.}~\bibnamefont {Seeley}}, \bibinfo {author} {\bibfnamefont {M.}~\bibnamefont {Kaicher}}, \bibinfo {author} {\bibfnamefont {J.}~\bibnamefont {McClean}}, \bibinfo {author} {\bibfnamefont {R.}~\bibnamefont {Babbush}}, \bibinfo {author} {\bibfnamefont {P.~V.}\ \bibnamefont {Coveney}}, \bibinfo {author} {\bibfnamefont {F.}~\bibnamefont {Mintert}}, \bibinfo {author} {\bibfnamefont {F.}~\bibnamefont {Wilhelm}},\ and\ \bibinfo {author} {\bibfnamefont {P.~J.}\ \bibnamefont {Love}},\ }\bibfield  {title} {\bibinfo {title} {{The Bravyi\textendash{}Kitaev transformation: Properties and applications}},\ }\href {https://doi.org/10.1002/qua.24969} {\bibfield  {journal} {\bibinfo  {journal} {Int. J. Quant. Chem.}\ }\textbf {\bibinfo {volume} {115}},\ \bibinfo {pages} {1431} (\bibinfo {year} {2015})}\BibitemShut {NoStop}%
\bibitem [{\citenamefont {Hastings}\ \emph {et~al.}(2015)\citenamefont {Hastings}, \citenamefont {Wecker}, \citenamefont {Bauer},\ and\ \citenamefont {Troyer}}]{hastings2015}%
  \BibitemOpen
  \bibfield  {author} {\bibinfo {author} {\bibfnamefont {M.~B.}\ \bibnamefont {Hastings}}, \bibinfo {author} {\bibfnamefont {D.}~\bibnamefont {Wecker}}, \bibinfo {author} {\bibfnamefont {B.}~\bibnamefont {Bauer}},\ and\ \bibinfo {author} {\bibfnamefont {M.}~\bibnamefont {Troyer}},\ }\bibfield  {title} {\bibinfo {title} {Improving quantum algorithms for quantum chemistry},\ }\href@noop {} {\bibfield  {journal} {\bibinfo  {journal} {Quantum Info. Comput.}\ }\textbf {\bibinfo {volume} {15}},\ \bibinfo {pages} {1–21} (\bibinfo {year} {2015})}\BibitemShut {NoStop}%
\bibitem [{\citenamefont {Whitfield}\ \emph {et~al.}(2016)\citenamefont {Whitfield}, \citenamefont {Havl\'\i{}\v{c}ek},\ and\ \citenamefont {Troyer}}]{Whitfield:2016lhw}%
  \BibitemOpen
  \bibfield  {author} {\bibinfo {author} {\bibfnamefont {J.~D.}\ \bibnamefont {Whitfield}}, \bibinfo {author} {\bibfnamefont {V.}~\bibnamefont {Havl\'\i{}\v{c}ek}},\ and\ \bibinfo {author} {\bibfnamefont {M.}~\bibnamefont {Troyer}},\ }\bibfield  {title} {\bibinfo {title} {{Local spin operators for fermion simulations}},\ }\href {https://doi.org/10.1103/PhysRevA.94.030301} {\bibfield  {journal} {\bibinfo  {journal} {Phys. Rev. A}\ }\textbf {\bibinfo {volume} {94}},\ \bibinfo {pages} {030301} (\bibinfo {year} {2016})}\BibitemShut {NoStop}%
\bibitem [{\citenamefont {Havlíček}\ \emph {et~al.}(2017)\citenamefont {Havlíček}, \citenamefont {Troyer},\ and\ \citenamefont {Whitfield}}]{Havl_ek_2017}%
  \BibitemOpen
  \bibfield  {author} {\bibinfo {author} {\bibfnamefont {V.}~\bibnamefont {Havlíček}}, \bibinfo {author} {\bibfnamefont {M.}~\bibnamefont {Troyer}},\ and\ \bibinfo {author} {\bibfnamefont {J.~D.}\ \bibnamefont {Whitfield}},\ }\bibfield  {title} {\bibinfo {title} {Operator locality in the quantum simulation of fermionic models},\ }\bibfield  {journal} {\bibinfo  {journal} {Physical Review A}\ }\textbf {\bibinfo {volume} {95}},\ \href {https://doi.org/10.1103/physreva.95.032332} {10.1103/physreva.95.032332} (\bibinfo {year} {2017})\BibitemShut {NoStop}%
\bibitem [{\citenamefont {Setia}\ \emph {et~al.}(2019)\citenamefont {Setia}, \citenamefont {Bravyi}, \citenamefont {Mezzacapo},\ and\ \citenamefont {Whitfield}}]{Setia_2019}%
  \BibitemOpen
  \bibfield  {author} {\bibinfo {author} {\bibfnamefont {K.}~\bibnamefont {Setia}}, \bibinfo {author} {\bibfnamefont {S.}~\bibnamefont {Bravyi}}, \bibinfo {author} {\bibfnamefont {A.}~\bibnamefont {Mezzacapo}},\ and\ \bibinfo {author} {\bibfnamefont {J.~D.}\ \bibnamefont {Whitfield}},\ }\bibfield  {title} {\bibinfo {title} {Superfast encodings for fermionic quantum simulation},\ }\bibfield  {journal} {\bibinfo  {journal} {Physical Review Research}\ }\textbf {\bibinfo {volume} {1}},\ \href {https://doi.org/10.1103/physrevresearch.1.033033} {10.1103/physrevresearch.1.033033} (\bibinfo {year} {2019})\BibitemShut {NoStop}%
\bibitem [{\citenamefont {Steudtner}\ and\ \citenamefont {Wehner}(2019)}]{steudtner_quantum_2019}%
  \BibitemOpen
  \bibfield  {author} {\bibinfo {author} {\bibfnamefont {M.}~\bibnamefont {Steudtner}}\ and\ \bibinfo {author} {\bibfnamefont {S.}~\bibnamefont {Wehner}},\ }\bibfield  {title} {\bibinfo {title} {Quantum codes for quantum simulation of {Fermions} on a square lattice of qubits},\ }\href {https://doi.org/10.1103/PhysRevA.99.022308} {\bibfield  {journal} {\bibinfo  {journal} {Physical Review A}\ }\textbf {\bibinfo {volume} {99}},\ \bibinfo {pages} {022308} (\bibinfo {year} {2019})},\ \bibinfo {note} {arXiv:1810.02681 [quant-ph]}\BibitemShut {NoStop}%
\bibitem [{\citenamefont {Jiang}\ \emph {et~al.}(2019)\citenamefont {Jiang}, \citenamefont {McClean}, \citenamefont {Babbush},\ and\ \citenamefont {Neven}}]{jiang2019}%
  \BibitemOpen
  \bibfield  {author} {\bibinfo {author} {\bibfnamefont {Z.}~\bibnamefont {Jiang}}, \bibinfo {author} {\bibfnamefont {J.}~\bibnamefont {McClean}}, \bibinfo {author} {\bibfnamefont {R.}~\bibnamefont {Babbush}},\ and\ \bibinfo {author} {\bibfnamefont {H.}~\bibnamefont {Neven}},\ }\bibfield  {title} {\bibinfo {title} {Majorana loop stabilizer codes for error mitigation in fermionic quantum simulations},\ }\href {https://doi.org/10.1103/PhysRevApplied.12.064041} {\bibfield  {journal} {\bibinfo  {journal} {Phys. Rev. Appl.}\ }\textbf {\bibinfo {volume} {12}},\ \bibinfo {pages} {064041} (\bibinfo {year} {2019})}\BibitemShut {NoStop}%
\bibitem [{\citenamefont {Backens}\ \emph {et~al.}(2019)\citenamefont {Backens}, \citenamefont {Shnirman},\ and\ \citenamefont {Makhlin}}]{Backens_2019}%
  \BibitemOpen
  \bibfield  {author} {\bibinfo {author} {\bibfnamefont {S.}~\bibnamefont {Backens}}, \bibinfo {author} {\bibfnamefont {A.}~\bibnamefont {Shnirman}},\ and\ \bibinfo {author} {\bibfnamefont {Y.}~\bibnamefont {Makhlin}},\ }\bibfield  {title} {\bibinfo {title} {Jordan–wigner transformations for tree structures},\ }\bibfield  {journal} {\bibinfo  {journal} {Scientific Reports}\ }\textbf {\bibinfo {volume} {9}},\ \href {https://doi.org/10.1038/s41598-018-38128-8} {10.1038/s41598-018-38128-8} (\bibinfo {year} {2019})\BibitemShut {NoStop}%
\bibitem [{\citenamefont {Jiang}\ \emph {et~al.}(2020)\citenamefont {Jiang}, \citenamefont {Kalev}, \citenamefont {Mruczkiewicz},\ and\ \citenamefont {Neven}}]{Jiang_2020}%
  \BibitemOpen
  \bibfield  {author} {\bibinfo {author} {\bibfnamefont {Z.}~\bibnamefont {Jiang}}, \bibinfo {author} {\bibfnamefont {A.}~\bibnamefont {Kalev}}, \bibinfo {author} {\bibfnamefont {W.}~\bibnamefont {Mruczkiewicz}},\ and\ \bibinfo {author} {\bibfnamefont {H.}~\bibnamefont {Neven}},\ }\bibfield  {title} {\bibinfo {title} {Optimal fermion-to-qubit mapping via ternary trees with applications to reduced quantum states learning},\ }\href {https://doi.org/10.22331/q-2020-06-04-276} {\bibfield  {journal} {\bibinfo  {journal} {Quantum}\ }\textbf {\bibinfo {volume} {4}},\ \bibinfo {pages} {276} (\bibinfo {year} {2020})}\BibitemShut {NoStop}%
\bibitem [{\citenamefont {Derby}\ \emph {et~al.}(2021)\citenamefont {Derby}, \citenamefont {Klassen}, \citenamefont {Bausch},\ and\ \citenamefont {Cubitt}}]{Derby_2021}%
  \BibitemOpen
  \bibfield  {author} {\bibinfo {author} {\bibfnamefont {C.}~\bibnamefont {Derby}}, \bibinfo {author} {\bibfnamefont {J.}~\bibnamefont {Klassen}}, \bibinfo {author} {\bibfnamefont {J.}~\bibnamefont {Bausch}},\ and\ \bibinfo {author} {\bibfnamefont {T.}~\bibnamefont {Cubitt}},\ }\bibfield  {title} {\bibinfo {title} {Compact fermion to qubit mappings},\ }\bibfield  {journal} {\bibinfo  {journal} {Physical Review B}\ }\textbf {\bibinfo {volume} {104}},\ \href {https://doi.org/10.1103/physrevb.104.035118} {10.1103/physrevb.104.035118} (\bibinfo {year} {2021})\BibitemShut {NoStop}%
\bibitem [{\citenamefont {Chiew}\ and\ \citenamefont {Strelchuk}(2023)}]{Chiew_2023}%
  \BibitemOpen
  \bibfield  {author} {\bibinfo {author} {\bibfnamefont {M.}~\bibnamefont {Chiew}}\ and\ \bibinfo {author} {\bibfnamefont {S.}~\bibnamefont {Strelchuk}},\ }\bibfield  {title} {\bibinfo {title} {Discovering optimal fermion-qubit mappings through algorithmic enumeration},\ }\href {https://doi.org/10.22331/q-2023-10-18-1145} {\bibfield  {journal} {\bibinfo  {journal} {Quantum}\ }\textbf {\bibinfo {volume} {7}},\ \bibinfo {pages} {1145} (\bibinfo {year} {2023})}\BibitemShut {NoStop}%
\bibitem [{\citenamefont {Miller}\ \emph {et~al.}(2023)\citenamefont {Miller}, \citenamefont {Zimborás}, \citenamefont {Knecht}, \citenamefont {Maniscalco},\ and\ \citenamefont {García-Pérez}}]{Miller_2023}%
  \BibitemOpen
  \bibfield  {author} {\bibinfo {author} {\bibfnamefont {A.}~\bibnamefont {Miller}}, \bibinfo {author} {\bibfnamefont {Z.}~\bibnamefont {Zimborás}}, \bibinfo {author} {\bibfnamefont {S.}~\bibnamefont {Knecht}}, \bibinfo {author} {\bibfnamefont {S.}~\bibnamefont {Maniscalco}},\ and\ \bibinfo {author} {\bibfnamefont {G.}~\bibnamefont {García-Pérez}},\ }\bibfield  {title} {\bibinfo {title} {Bonsai algorithm: Grow your own fermion-to-qubit mappings},\ }\bibfield  {journal} {\bibinfo  {journal} {PRX Quantum}\ }\textbf {\bibinfo {volume} {4}},\ \href {https://doi.org/10.1103/prxquantum.4.030314} {10.1103/prxquantum.4.030314} (\bibinfo {year} {2023})\BibitemShut {NoStop}%
\bibitem [{\citenamefont {Chen}\ and\ \citenamefont {Xu}(2023)}]{chen_equivalence_2023}%
  \BibitemOpen
  \bibfield  {author} {\bibinfo {author} {\bibfnamefont {Y.-A.}\ \bibnamefont {Chen}}\ and\ \bibinfo {author} {\bibfnamefont {Y.}~\bibnamefont {Xu}},\ }\bibfield  {title} {\bibinfo {title} {Equivalence between {Fermion}-to-{Qubit} {Mappings} in two {Spatial} {Dimensions}},\ }\href {https://doi.org/10.1103/PRXQuantum.4.010326} {\bibfield  {journal} {\bibinfo  {journal} {PRX Quantum}\ }\textbf {\bibinfo {volume} {4}},\ \bibinfo {pages} {010326} (\bibinfo {year} {2023})},\ \bibinfo {note} {publisher: American Physical Society}\BibitemShut {NoStop}%
\bibitem [{\citenamefont {O'Brien}\ and\ \citenamefont {Strelchuk}(2024)}]{PhysRevB.109.115149}%
  \BibitemOpen
  \bibfield  {author} {\bibinfo {author} {\bibfnamefont {O.}~\bibnamefont {O'Brien}}\ and\ \bibinfo {author} {\bibfnamefont {S.}~\bibnamefont {Strelchuk}},\ }\bibfield  {title} {\bibinfo {title} {Ultrafast hybrid fermion-to-qubit mapping},\ }\href {https://doi.org/10.1103/PhysRevB.109.115149} {\bibfield  {journal} {\bibinfo  {journal} {Phys. Rev. B}\ }\textbf {\bibinfo {volume} {109}},\ \bibinfo {pages} {115149} (\bibinfo {year} {2024})}\BibitemShut {NoStop}%
\bibitem [{\citenamefont {Chen}\ \emph {et~al.}(2018)\citenamefont {Chen}, \citenamefont {Kapustin},\ and\ \citenamefont {Radicevic}}]{Chen_2018}%
  \BibitemOpen
  \bibfield  {author} {\bibinfo {author} {\bibfnamefont {Y.-A.}\ \bibnamefont {Chen}}, \bibinfo {author} {\bibfnamefont {A.}~\bibnamefont {Kapustin}},\ and\ \bibinfo {author} {\bibfnamefont {D.}~\bibnamefont {Radicevic}},\ }\bibfield  {title} {\bibinfo {title} {Exact bosonization in two spatial dimensions and a new class of lattice gauge theories},\ }\href {https://doi.org/10.1016/j.aop.2018.03.024} {\bibfield  {journal} {\bibinfo  {journal} {Annals of Physics}\ }\textbf {\bibinfo {volume} {393}},\ \bibinfo {pages} {234–253} (\bibinfo {year} {2018})}\BibitemShut {NoStop}%
\bibitem [{\citenamefont {Zohar}\ and\ \citenamefont {Cirac}(2018)}]{Zohar_2018}%
  \BibitemOpen
  \bibfield  {author} {\bibinfo {author} {\bibfnamefont {E.}~\bibnamefont {Zohar}}\ and\ \bibinfo {author} {\bibfnamefont {J.~I.}\ \bibnamefont {Cirac}},\ }\bibfield  {title} {\bibinfo {title} {Eliminating fermionic matter fields in lattice gauge theories},\ }\bibfield  {journal} {\bibinfo  {journal} {Physical Review B}\ }\textbf {\bibinfo {volume} {98}},\ \href {https://doi.org/10.1103/physrevb.98.075119} {10.1103/physrevb.98.075119} (\bibinfo {year} {2018})\BibitemShut {NoStop}%
\bibitem [{\citenamefont {Bultinck}\ \emph {et~al.}(2017)\citenamefont {Bultinck}, \citenamefont {Williamson}, \citenamefont {Haegeman},\ and\ \citenamefont {Verstraete}}]{Bultinck_2017}%
  \BibitemOpen
  \bibfield  {author} {\bibinfo {author} {\bibfnamefont {N.}~\bibnamefont {Bultinck}}, \bibinfo {author} {\bibfnamefont {D.~J.}\ \bibnamefont {Williamson}}, \bibinfo {author} {\bibfnamefont {J.}~\bibnamefont {Haegeman}},\ and\ \bibinfo {author} {\bibfnamefont {F.}~\bibnamefont {Verstraete}},\ }\bibfield  {title} {\bibinfo {title} {Fermionic matrix product states and one-dimensional topological phases},\ }\bibfield  {journal} {\bibinfo  {journal} {Physical Review B}\ }\textbf {\bibinfo {volume} {95}},\ \href {https://doi.org/10.1103/physrevb.95.075108} {10.1103/physrevb.95.075108} (\bibinfo {year} {2017})\BibitemShut {NoStop}%
\bibitem [{\citenamefont {Cirac}\ \emph {et~al.}(2021)\citenamefont {Cirac}, \citenamefont {Pérez-García}, \citenamefont {Schuch},\ and\ \citenamefont {Verstraete}}]{Cirac_2021}%
  \BibitemOpen
  \bibfield  {author} {\bibinfo {author} {\bibfnamefont {J.~I.}\ \bibnamefont {Cirac}}, \bibinfo {author} {\bibfnamefont {D.}~\bibnamefont {Pérez-García}}, \bibinfo {author} {\bibfnamefont {N.}~\bibnamefont {Schuch}},\ and\ \bibinfo {author} {\bibfnamefont {F.}~\bibnamefont {Verstraete}},\ }\bibfield  {title} {\bibinfo {title} {Matrix product states and projected entangled pair states: Concepts, symmetries, theorems},\ }\bibfield  {journal} {\bibinfo  {journal} {Reviews of Modern Physics}\ }\textbf {\bibinfo {volume} {93}},\ \href {https://doi.org/10.1103/revmodphys.93.045003} {10.1103/revmodphys.93.045003} (\bibinfo {year} {2021})\BibitemShut {NoStop}%
\bibitem [{\citenamefont {Haegeman}\ \emph {et~al.}(2015)\citenamefont {Haegeman}, \citenamefont {Van~Acoleyen}, \citenamefont {Schuch}, \citenamefont {Cirac},\ and\ \citenamefont {Verstraete}}]{Haegeman_2015}%
  \BibitemOpen
  \bibfield  {author} {\bibinfo {author} {\bibfnamefont {J.}~\bibnamefont {Haegeman}}, \bibinfo {author} {\bibfnamefont {K.}~\bibnamefont {Van~Acoleyen}}, \bibinfo {author} {\bibfnamefont {N.}~\bibnamefont {Schuch}}, \bibinfo {author} {\bibfnamefont {J.~I.}\ \bibnamefont {Cirac}},\ and\ \bibinfo {author} {\bibfnamefont {F.}~\bibnamefont {Verstraete}},\ }\bibfield  {title} {\bibinfo {title} {Gauging quantum states: From global to local symmetries in many-body systems},\ }\bibfield  {journal} {\bibinfo  {journal} {Physical Review X}\ }\textbf {\bibinfo {volume} {5}},\ \href {https://doi.org/10.1103/physrevx.5.011024} {10.1103/physrevx.5.011024} (\bibinfo {year} {2015})\BibitemShut {NoStop}%
\bibitem [{\citenamefont {Xu}\ \emph {et~al.}(2023)\citenamefont {Xu}, \citenamefont {Chung}, \citenamefont {Qin}, \citenamefont {Schollwöck}, \citenamefont {White},\ and\ \citenamefont {Zhang}}]{xu2023coexistence}%
  \BibitemOpen
  \bibfield  {author} {\bibinfo {author} {\bibfnamefont {H.}~\bibnamefont {Xu}}, \bibinfo {author} {\bibfnamefont {C.-M.}\ \bibnamefont {Chung}}, \bibinfo {author} {\bibfnamefont {M.}~\bibnamefont {Qin}}, \bibinfo {author} {\bibfnamefont {U.}~\bibnamefont {Schollwöck}}, \bibinfo {author} {\bibfnamefont {S.~R.}\ \bibnamefont {White}},\ and\ \bibinfo {author} {\bibfnamefont {S.}~\bibnamefont {Zhang}},\ }\href@noop {} {\bibinfo {title} {Coexistence of superconductivity with partially filled stripes in the hubbard model}} (\bibinfo {year} {2023}),\ \Eprint {https://arxiv.org/abs/2303.08376} {arXiv:2303.08376 [cond-mat.supr-con]} \BibitemShut {NoStop}%
\bibitem [{\citenamefont {Lootens}\ \emph {et~al.}(2023{\natexlab{a}})\citenamefont {Lootens}, \citenamefont {Delcamp}, \citenamefont {Williamson},\ and\ \citenamefont {Verstraete}}]{lootens2023lowdepth}%
  \BibitemOpen
  \bibfield  {author} {\bibinfo {author} {\bibfnamefont {L.}~\bibnamefont {Lootens}}, \bibinfo {author} {\bibfnamefont {C.}~\bibnamefont {Delcamp}}, \bibinfo {author} {\bibfnamefont {D.}~\bibnamefont {Williamson}},\ and\ \bibinfo {author} {\bibfnamefont {F.}~\bibnamefont {Verstraete}},\ }\href@noop {} {\bibinfo {title} {Low-depth unitary quantum circuits for dualities in one-dimensional quantum lattice models}} (\bibinfo {year} {2023}{\natexlab{a}}),\ \Eprint {https://arxiv.org/abs/2311.01439} {arXiv:2311.01439 [quant-ph]} \BibitemShut {NoStop}%
\bibitem [{\citenamefont {Lootens}\ \emph {et~al.}(2023{\natexlab{b}})\citenamefont {Lootens}, \citenamefont {Delcamp}, \citenamefont {Ortiz},\ and\ \citenamefont {Verstraete}}]{Lootens_2023}%
  \BibitemOpen
  \bibfield  {author} {\bibinfo {author} {\bibfnamefont {L.}~\bibnamefont {Lootens}}, \bibinfo {author} {\bibfnamefont {C.}~\bibnamefont {Delcamp}}, \bibinfo {author} {\bibfnamefont {G.}~\bibnamefont {Ortiz}},\ and\ \bibinfo {author} {\bibfnamefont {F.}~\bibnamefont {Verstraete}},\ }\bibfield  {title} {\bibinfo {title} {Dualities in one-dimensional quantum lattice models: Symmetric hamiltonians and matrix product operator intertwiners},\ }\bibfield  {journal} {\bibinfo  {journal} {PRX Quantum}\ }\textbf {\bibinfo {volume} {4}},\ \href {https://doi.org/10.1103/prxquantum.4.020357} {10.1103/prxquantum.4.020357} (\bibinfo {year} {2023}{\natexlab{b}})\BibitemShut {NoStop}%
\bibitem [{\citenamefont {Lootens}\ \emph {et~al.}(2024)\citenamefont {Lootens}, \citenamefont {Delcamp},\ and\ \citenamefont {Verstraete}}]{Lootens:2022avn}%
  \BibitemOpen
  \bibfield  {author} {\bibinfo {author} {\bibfnamefont {L.}~\bibnamefont {Lootens}}, \bibinfo {author} {\bibfnamefont {C.}~\bibnamefont {Delcamp}},\ and\ \bibinfo {author} {\bibfnamefont {F.}~\bibnamefont {Verstraete}},\ }\bibfield  {title} {\bibinfo {title} {{Dualities in One-Dimensional Quantum Lattice Models: Topological Sectors}},\ }\href {https://doi.org/10.1103/PRXQuantum.5.010338} {\bibfield  {journal} {\bibinfo  {journal} {PRX Quantum}\ }\textbf {\bibinfo {volume} {5}},\ \bibinfo {pages} {010338} (\bibinfo {year} {2024})},\ \Eprint {https://arxiv.org/abs/2211.03777} {arXiv:2211.03777 [quant-ph]} \BibitemShut {NoStop}%
\end{thebibliography}%

\end{document}